\documentstyle[12pt,epsf]{article}

\newcommand{\eq}{\begin{equation}}
\newcommand{\en}{\end{equation}}
\newcommand{\be}{\begin{equation}}
\newcommand{\ee}{\end{equation}}
\newcommand{\eqa}{\begin{eqnarray}}
\newcommand{\ena}{\end{eqnarray}}
\newcommand{\ba}{\begin{eqnarray}}
\newcommand{\ea}{\end{eqnarray}}

\newcommand{\ZZ}{\hbox{{\rm Z{\hbox to 3pt{\hss\rm Z}}}}}

\newcommand{\half}{\mbox{\small$\frac{1}{2}$}}
\newcommand{\mye}{\mbox{e}}

\begin{document}
\begin{titlepage}
\vskip0.5cm
\vskip0.5cm
\begin{center}
{\Large\bf
The
two dimensional XY model at the transition temperature: \\ 
A high precision Monte Carlo study
}
\end{center}

\centerline{
Martin Hasenbusch
}
\vskip 0.4cm
\centerline{\sl 
Dipartimento di Fisica dell'
 Universit\`a di Pisa and INFN,}
\centerline{\sl Largo Bruno Pontecorvo 3,
I-56127 Pisa, Italy}
\vskip 0.3cm
\centerline{e-mail: Martin.Hasenbusch@df.unipi.it}
\vskip 0.4cm
\begin{abstract}
We study the classical XY (plane rotator) model at the Kosterlitz-Thouless
phase transition. We simulate the model using the single cluster algorithm 
on square lattices of a linear size up to $L=2048$. 
We derive the finite size behaviour of the second moment
correlation length over the lattice size $\xi_{2nd}/L$ at the transition 
temperature. This new prediction and the analogous one for the 
helicity modulus $\Upsilon$ are confronted with our Monte Carlo data.
This way $\beta_{KT}=1.1199$ is confirmed as inverse transition temperature.
Finally we address the puzzle of logarithmic corrections of the magnetic 
susceptibility $\chi$ at the transition temperature.
\end{abstract}

\end{titlepage}

\section{Introduction}
We study the classical XY model on the square lattice.
It is characterised by the action
\begin{equation}
\label{XYaction}
 S = - \beta \sum_{x,\mu}  \vec{s}_x  \vec{s}_{x + \hat \mu} \;\;,
\end{equation}
where $\vec{s}_x$ is a unit vector with two real components, 
 $x=(x_1,x_2)$ labels the sites
on the square lattice, 
where $x_1 \in \{1,2,...,L_1\}$ and $x_2 \in \{1,2,...,L_2\}$
\footnote{In our simulations
we use $L_1=L_2=L$ throughout}, 
$\mu$ gives the direction on
the lattice and $\hat \mu$ is a unit-vector in the $\mu$-direction. 
We consider periodic boundary conditions in both directions.
The coupling constant has been set 
to $J=1$ and $\beta$ is the inverse temperature.
In our 
notation, the Boltzmann-factor is given by $\exp(-S)$. Sometimes 
in the literature the present model is also called ``plane rotator model'',
while the name XY-model is used for a model with three spin-components.

Kosterlitz and Thouless \cite{KT} have argued that the XY-model 
undergoes a phase 
transition of infinite order. The low temperature phase is characterised
by a vanishing order parameter 
and an infinite correlation length $\xi$, associated with a line of Gaussian 
fixed points. At a sufficiently high temperature pairs of vortices unbind and
start to disorder the 
system resulting in a finite correlation length $\xi$. 
In the neighbourhood of the 
transition temperature $T_{KT}$ it behaves as
\begin{equation}
\label{xiKT}
\xi \simeq a \exp(b \; t^{-1/2}) \;\;,
\end{equation}
where $t=(T-T_{KT})/T_{KT}$ is the reduced temperature and $a$ and $b$ are 
non-universal constants.
In subsequent work  (e.g. refs. \cite{Jo77,AmGoGr80}) the
results of Kosterlitz and Thouless had been 
confirmed and the arguments had been put on a more rigorous basis.

This rather good theoretical understanding of the  Kosterlitz-Thouless (KT)
phase transition is contrasted by the fact that the verification 
of the theoretical predictions in Monte Carlo simulations had often 
been inconclusive or even in contradiction. 
Only starting from the early nineties, Monte Carlo simulations allowed to
favour clearly the KT-behaviour~(\ref{xiKT}) over a power law
$\xi\propto t^{-\nu}$, which is characteristic for a second order    
phase transition. A typical example for such a work is ref. \cite{JaNa93},
where the XY model with the Villian action \cite{villain} was studied 
on lattices of a size up to $1200^2$. 

The difficulties in Monte Carlo simulations
might be explained by logarithmic corrections that are predicted to be 
present in the neighbourhood of the transition.

In the present paper we like to address two puzzling results presented in the 
literature that are related to this problem: \\
\begin{itemize}
\item
The two most precise results \cite{olsson,HaPi97} 
for the transition temperature $T_{KT}$
of the XY-model differ by about 8 times the quoted errors.  
\item
The magnetic susceptibility is predicted to scale as 
$\chi \propto L^{2-\eta} (\ln L)^{-2 r}$ with $\eta=1/4$ and $r=-1/16$
at the transition temperature. \footnote{Note that the analogous result
$\chi \propto \xi^{2-\eta} (\ln \xi)^{-2 r}$  
for the thermodynamic limit 
in the high temperature phase does not hold. In refs. \cite{Ba00,Ba01}
it was argued and numerically verified that 
instead $\chi \propto \xi^{2-\eta} (1 + c/(\ln \xi + u)^2+...)$ is correct. }
 However the authors of 
refs. \cite{IrKe95,Janke96} 
find in their Monte Carlo simulations $r = - 0.023(10)$
\footnote{The authors confirmed their numerical result
for $r$  by a study of Lee-Yang zeros \cite{IrKe96}}
and $r = - 0.0270(10)$,
respectively. 
\end{itemize}

In refs. \cite{HaMaPi94,HaPi97} the authors have shown that XY models with 
different actions share the universality class of the BCSOS model. 
This had been achieved by matching the renormalization group (RG) 
flow of the BCSOS model at the 
critical point with that of the exact duals \cite{savit} 
of the XY models using a
particular Monte Carlo renormalization group method. 
As a result of this matching  
the estimate $\beta_{KT}=1.1199(1)=1/0.89294(8)$
for the XY model~(\ref{XYaction}) has been obtained.
\footnote{In the case of the Villian action, the matching method gives 
$\beta_{V,KT}=0.7515(2)$, while the authors of ref. \cite{JaNa93} had found
$\beta_{V,KT}=0.752(5)$ fitting their data for the correlation length with the 
ansatz~(\ref{xiKT}) and a similar fit for the magnetic susceptibility.}
The BCSOS model is equivalent with the six-vertex model \cite{beijeren77}.
The exact 
result for the correlation length of the six-vertex model \cite{lieb,wu,baxter}
shows the behaviour of  eq.~(\ref{xiKT}) predicted by the KT-theory. 
The main advantage of the matching approach is that the logarithmic
corrections and in particular also subleading logarithmic corrections
are the same in the XY-model and the BCSOS model. 
\footnote{A brief discussion of this fact will be
 given in section \ref{KTtheory}.}

In a more standard approach, 
Olsson \cite{olsson} and
Schultka and Manousakis \cite{ScMa94} have 
studied the finite size behaviour of  the helicity modulus arriving at 
the estimates $1/\beta_{KT}=0.89213(10)$ and $1/\beta_{KT}=0.89220(13)$, 
respectively. These authors studied lattice sizes up to $L=256$ and $L=400$,
respectively. While in their approach leading logarithmic corrections are 
taken properly into account, subleading logarithmic corrections are missed.
This might explain the missmatch of the results for the transition 
temperature.
Here we shall resolve this discrepancy by brute force: We study
the helicity modulus (and in addition the second moment correlation length)
on lattices up to $L=2048$. 

Having an accurate estimate of $T_{KT}$ and numerical results for large 
lattice sizes at hand, we then study 
the scaling of the magnetic susceptibility. 
Here it turns out that the puzzling result for the value of the exponent $r$ 
can be resolved by taking into account subleading corrections.

A major purpose of the present paper is to check the reliability of standard
methods to determine the temperature of the transition
and to verify its KT-nature. 
This aims mainly at more complicated models,
e.g. quantum models or thin films of 
three dimensional systems with nontrivial boundary conditions,
where the duality transformation is not possible, and hence the method
of refs. \cite{HaMaPi94,HaPi97} can not be applied.

The outline of the paper is the following: In the next section we give the 
definitions of the observables that are studied in this paper: the 
helicity modulus, the second moment correlation length and the magnetic 
susceptibility. Next we summarise some results from the literature 
on duality and the  
RG-flow at the KT-transition. 
We re-derive  the finite size behaviour of the helicity modulus at the 
transition temperature. Along the same lines we then derive a new result
for the dimensionless ratio $\xi_{2nd}/L$.  
This is followed by Monte Carlo simulations using the single cluster algorithm 
for lattices of a linear size up to $L=2048$ for $\beta=1.1199$ and 
$\beta=1.12091$. Fitting the data for $\beta=1.1199$ we find the behaviour 
of the helicity modulus and $\xi_{2nd}/L$ predicted by the theory for the 
transition temperature, while for $\beta=1.12091$ there is clear missmatch.
Finally we analyse the data of the magnetic susceptibility at $\beta=1.1199$.

\section{The observables}
In this section we shall summarise the definitions of the observables 
that we have measured in our simulations.
The total magnetisation is defined by
\begin{equation}
 \vec{M} =  \sum_x \vec{s}_x \;\;.
\end{equation}
The magnetic susceptibility is then given as 
\begin{equation}
 \chi =  \frac{1}{L^2} \vec{M}^2 \;\;.
\end{equation}

\subsection{The second moment correlation length $\xi_{2nd}$}
The second moment correlation length on a lattice of the 
size $L^2$  is defined by
\begin{equation}
\label{second}
\xi_{2nd}=\frac{1}{2 \sin(\pi/L)} \left(\frac{\chi}{F}-1\right)^{1/2} \;\;,
\end{equation}
where  $\chi$ is the magnetic susceptibility as defined above
and
\begin{equation}
 F =\frac{1}{L^2} 
\sum_{x,y} \langle \vec{s}_x \vec{s}_y \rangle  \cos(2 \pi (y_1-x_1)/L) \;\;.
\end{equation}
Note that the results obtained in this paper only hold for the definition
of $\xi_{2nd}$ given in this subsection.

\subsection{The helicity modulus $\Upsilon$}
 The helicity modulus $\Upsilon$ gives the reaction of the system under
 a torsion \cite{helidef}.
 To define the helicity modulus we consider a system, where
 rotated boundary conditions in one direction are introduced:
 For pairs $x,y$ of nearest neighbour sites on the lattice with 
 $x_1=L_1$, $y_1=1$ and $x_2=y_2$ the term $\vec{s}_x  \vec{s}_y$
 is replaced by
\begin{equation}
 \vec{s_x} \cdot R_{\alpha} \vec{s_y} =
 s_x^{(1)} \left(\cos(\alpha) s_x^{(1)} 
+ \sin(\alpha) s_x^{(2)} \right)+ s_x^{(2)}\left(
\cos(\alpha) s_x^{(2)}
-\sin(\alpha) s_x^{(1)}\right) \;\;.
\end{equation}
The helicity modulus  
is then defined by the second derivative of the
free energy with respect to $\alpha$ at $\alpha=0$
\begin{equation}
\label{ups_inf}
\Upsilon = -  \frac{L_1}{L_2}
\left .
\frac{\partial^2 \ln Z(\alpha)}{\partial \alpha^2} \right |_{\alpha=0} \;\;.
\end{equation}
Note that we have skipped a factor one over temperature in our definition 
of the helicity modulus to obtain a dimensionless quantity. 
It is easy to write the helicity modulus as an observable of the 
system at $\alpha=0$ \cite{heliform}. For $L_1=L_2=L$ we get
\begin{equation}
\Upsilon = \frac{\beta}{L^2} 
\left \langle \vec{s}_{x \phantom{\hat 1} } \vec{s}_{x+\hat 1} \right \rangle
-  \frac{\beta^2}{L^2} \left \langle \left(
  s_{x\phantom{\hat 1}}^{(1)} s_{x+\hat 1}^{(2)} 
- s_{x\phantom{\hat 1}}^{(2)} s_{x+\hat 1}^{(1)} 
\right)^2 \right \rangle \;\;.
\end{equation}

\section{KT-theory}
\label{KTtheory}
In this section we summarise results from the literature that are
relevant for our numerical study and also derive a novel result for the 
finite size behaviour of the second moment correlation length at the 
transition temperature.

XY models can be exactly mapped by
a so called duality transformation \cite{savit} into solid on solid (SOS) 
models. E.g. the XY model with the action~(\ref{XYaction}) becomes
\be
Z_{XY}^{SOS} = \sum_{\{h\}} \prod_{x,\mu} I_{|h_x - h_{x+\hat \mu}|}(\beta) \, ,
\ee
where the $I_n$ are modified Bessel functions and the $h_x$ are integer.
The XY model with Villian action \cite{villain} takes a simpler form under duality:
\be
 Z_{V}^{SOS}  = \sum_{\{h\}} 
\exp\left(-\frac{1}{2 \beta} \sum_{x,\mu}
(h_x - h_{x+\hat \mu})^2 \right) \;\; ,
\ee
where the $h_x$ are integer again. This model is also called discrete Gaussian
(DG) model. In the context of finite size scaling one
should pay attention to the fact that the boundary conditions transform 
non-trivially under duality. E.g. periodic boundary conditions in the XY model 
require that in the SOS model one sums over all integer shifts $h_1$ and 
$h_2$ at the boundaries in $1$- and $2$-direction, respectively.

It turned out to be most convenient to study the Kosterlitz-Thouless
phase transition using generalisations of SOS models
(see e.g. refs. \cite{Jo77,AmGoGr80}).

\subsection{The Sine-Gordon model}
The Sine-Gordon model is defined by the action
\begin{equation}
\label{sgaction}
S_{SG} = \frac{1}{2 \beta} \sum_{x,\mu} (\phi_x - \phi_{x+\hat \mu})^2
 - z \sum_x  \cos(2 \pi \phi_x)  \;\;,
\end{equation}
where the variables $\phi_x$ are real numbers. For positive values of $z$, 
the periodic potential favours $\phi_x$ close to integers. In particular, 
in the limit $z \rightarrow \infty$, we recover the DG-SOS model.
In the limit $z=0$ we get the Gaussian model (or in the language of 
high energy physics, a free field theory).   The Sine-Gordon model 
(using cutoff schemes different from the lattice) 
can be used to derive the RG-flow associated with the KT phase transition.
For $\beta> 2/\pi$ the coupling $z$ is irrelevant, while for $\beta < 2/\pi$ 
it becomes relevant. 
To discuss the RG-flow it is convenient to  define
\begin{equation}
x=\pi \beta -2 \;\;.
\end{equation}
The flow-equations are derived in the neighbourhood of $(x,z)=(0,0)$.
To leading order they are given by
\begin{eqnarray}
  \frac{\partial z}{\partial t}  &=&  -x z \; + \; ... \;\;, \\
  \frac{\partial x}{\partial t}  &=&  - const \;\; z^2  \;  + \; ...\;\;,
\end{eqnarray}
where $t=\ln l$ is the logarithm of the length scale $l$ at which the coupling 
is taken. Note that we consider a fixed lattice spacing and a running length
scale $l$, while e.g. in ref. \cite{AmGoGr80} the cutoff scale is varied. 
This explains the opposite sign in the flow equations compared with e.g. 
ref. \cite{AmGoGr80}. 
The $const$ in the equation above depends on the particular type 
of cut-off that is used.  Corrections of $O(z^3)$ 
 have been computed in ref. \cite{AmGoGr80} and confirmed in ref.
\cite{BaHe00}.
Here we are mainly interested in the finite size behaviour at the transition 
temperature. Therefore the trajectory at the transition temperature is of 
particular interest. It is characterised by the fact that it ends in 
$(x,z)=(0,0)$.  To leading order it is given by
\be
 x= const^{1/2} \;\; z  \;\;.
\ee
It follows that the RG-flow on the critical trajectory is given by
\be
 \frac{\partial x}{\partial t} = - x^2 \;\;.
\ee
I.e. on the critical trajectory
\be
\label{xtransition}
 x = \frac{1}{\ln l + C} \;\;,
\ee
where $C$ is an integration constant that depends on the initial value $x_i$ 
of $x$ at $l=1$. 
Taking into account the  next to leading order result of ref. \cite{AmGoGr80}
the flow on the critical trajectory becomes
\be
  \frac{\partial x}{\partial t} = - x^2 - \frac{1}{2} x^3 \; ... \;.
\ee
Implicitly the solution is given by \cite{AmGoGr80}
\be
\label{mainamit}
 \ln l = \frac{1}{x} - \frac{1}{x_i}
                     - \frac{1}{2} \ln \frac{1/x+1/2}{1/x_i+1/2} \;\;,
\ee
where now the initial value $x_i$  of $x$
takes the role of the integration constant. The authors of ref. \cite{AmGoGr80}
give an approximate solution of this equation that is valid for $x_i >> x$. 
This leads to corrections to eq.~(\ref{xtransition}) that are 
proportional to  $\ln |\ln L|/|\ln L|^2$. However, 
in our numerical simulations we are rather in a situation where $x_i$ and $x$
differ only by a small factor. Therefore we make no attempt to fit our data
taking explicitly into account the last term of eq.~(\ref{mainamit}).

An important result of ref. \cite{AmGoGr80} is that corrections 
proportional to \\
$\ln |\ln L|/|\ln L|^2$ arise from the RG-flow in the $(x,z)$-plane and are
not caused by some 
additional marginal operators, which might have different amplitudes in 
different models. Therefore the two-parameter matching of refs. 
\cite{HaPi97,HaMaPi94} is sufficient to take properly into account
corrections proportional to  $\ln |\ln L|/|\ln L|^2$ (and beyond).

\subsection{Finite size scaling of dimensionless quantities}
Here we compute the values of the helicity modulus $\Upsilon$ and the 
ratio $\xi_{2nd}/L$  at $T_{KT}$ in the limit $L\rightarrow \infty$ and 
leading $1/\ln L$ corrections to it. 
Since for both quantities the coefficient of the order $z$ is vanishing, 
this can be achieved by computing both quantities at $z=0$ (i.e. for the 
Gaussian model) and plugging in 
the value of $\beta$ given by eq.~(\ref{xtransition}).

\subsubsection{The helicity modulus}
The helicity modulus can be easily expressed in terms of the SOS model
dual to the XY model:
\begin{equation}
\Upsilon= \frac{L_2}{L_1} \langle h_1^2 \rangle_{SOS} \;\;,
\end{equation}
where $h_1$ is the shift at the boundary in the $1$-direction. In this 
form we can compute the helicity modulus in the Sine-Gordon model. 
To
this end we have to compute the free energy as a function of the boundary shifts
$h_1,h_2$:
\begin{equation}
 F(h_1,h_2) = - \ln(Z(h_1,h_2)/Z(0,0)) \;\;,
\end{equation}
where $Z(h_1,h_2)$ is the partition function of the system with 
a shift by $h_1$ and 
$h_2$ at the boundaries in $1$ and $2$-direction, respectively.
From the SG-action~(\ref{sgaction}) we directly read off that $F(h_1,h_2)$
is an even function of $z$. Hence the leading $z$ dependent contribution is
$O(z^2)$. Hence for our purpose the purely Gaussian result $z=0$
is sufficient.
For the action ~(\ref{sgaction}) at $z=0$ we get
\begin{eqnarray}
Z(h_1,h_2) &=&\int \mbox{D} [\phi] 
\;\; \exp\left(-\frac{1}{2 \beta} \sum_{x,\mu} 
             (\phi_x - \phi_{x+\hat \mu}-d_{\mu})^2\right) \nonumber \\
 &=& \int \mbox{D} [\phi] \;\;
 \exp\left(- \frac{1}{2 \beta}[L_1 L_2 (d_{1}^2+d_{2}^2) + \sum_{x,\mu} 
(\phi_x - \phi_{x+\hat \mu})^2 ] 
 \right) \nonumber \\
&=& \exp\left(- \frac{1}{2 \beta} L_1 L_2 (d_{1}^2+d_{2}^2) \right)\;\; Z(0,0)
\nonumber \\
&=& 
\exp\left(- \frac{1}{2 \beta} 
\left[ \frac{L_2}{L_1} h_{1}^2 + \frac{L_1}{L_2} h_{2}^2
\right] \right)\;\; Z(0,0) \;\;,
\end{eqnarray}
where we have defined $d_{\mu}=h_{\mu}/L_{\mu}$. Note that we have distributed 
the boundary shift along the lattice by a reparametrisation of the field:
\begin{equation}
\label{shift}
\phi_x = \tilde \phi_x + x_1 d_1 + x_2 d_2 \;\;,
\end{equation}
where $\tilde \phi_x$ is the original field.
It follows 
\be
\label{helires1}
\Upsilon = \frac{L_2}{L_1} \frac{\sum_{h_1}
\exp\left(- \frac{1}{2 \beta} \frac{L_2}{L_1} h_{1}^2 \right) \;\; h_1^2
    }
 {\sum_{h_1} \exp\left(- \frac{1}{2 \beta} \frac{L_2}{L_1} h_{1}^2 \right) }
\;\;.
\ee

Alternatively we might evaluate the helicity modulus in the spin-wave limit
of the XY model on the original lattice. This is justified by the duality
transformation presented in ref. \cite{Jo77} in appendix D. Here we are
only interested in the Gaussian limit of the model. Under duality the $\beta$
of the Gaussian model transforms as $\tilde \beta = 1/\beta$.  Secondly we
have to take into account that even though vortices are not present in
the limit $z=0$, the periodicity of the XY model has to be taken into account 
for the boundary conditions. Hence, 
the proper spin-wave (SW) description of the 
XY-model on a finite lattice with periodic boundary conditions is 
\begin{equation}
\label{winding}
 Z_{SW} = \sum_{n_1,n_2}    
W(n_1,n_2) Z(0,0) \;\;,
\end{equation}
where $n_1$ and $n_2$ count the windings of the XY-field along the $1$ and $2$
direction respectively. In the Gaussian model they are given by shifts by 
$2 \pi n_1$ and $2 \pi n_2$ at the boundaries.
The corresponding weights are
\begin{equation}
\label{weight}
W(n_1,n_2) = \exp\left(- \frac{ (2 \pi)^2 }{2 \tilde \beta}
\left[ \frac{L_2}{L_1} n_{1}^2 + \frac{L_1}{L_2} n_{2}^2 \right] \right) \;\;.
\end{equation}
Here we can easily introduce a rotation by the angle $\alpha$ at the boundary:
\begin{equation}
 Z_{SW,\alpha} = \sum_{n_1,n_2}   \exp\left(- \frac{(2 \pi)^2 }{2 \tilde \beta}
\left[ \frac{L_2}{L_1} [n_{1}+\alpha/(2 \pi)]^2 + \frac{L_1}{L_2} n_{2}^2
\right] \right) Z(0,0) \;\;.
\end{equation}
 Plugging this result into the definition~(\ref{ups_inf}) 
of the helicity modulus we get
\begin{equation}
\label{helires2}
\Upsilon =
\frac{1}{\tilde \beta} - \frac{L_2}{L_1} 
\frac{ 
\sum_{n_1} 
\exp\left(
-\frac{(2 \pi n_1)^2 }{2 \tilde \beta}  \frac{L_2}{L_1} 
\right)  \left[\frac{2 \pi n_1}{\tilde \beta} \frac{L_2}{L_1}\right]^2
}
{ \sum_{n_1} 
\exp\left(
-\frac{(2 \pi n_1)^2 }{2  \tilde \beta}  \frac{L_2}{L_1} 
\right) } \;\;.
\end{equation}
In the literature often only $\Upsilon=1/\tilde \beta =\beta$ is 
quoted and the (tiny) correction due to  winding fields is ignored.
We have checked numerically that the results of eq.~(\ref{helires1})
and eq.~(\ref{helires2}) indeed coincide. Here we are interested in the 
case of an $L^2$ lattice in the neighbourhood of $\beta=2/\pi$. 
One gets
\begin{equation}
\Upsilon_{L^2,z=0}=0.63650817819... + 1.001852182... \; (\beta-2/\pi) 
 + ... \;\;.
\end{equation}
Plugging in the result~(\ref{xtransition}) and identifying the lattice size
$L$ with the scale at which the coupling is taken, we get
\begin{equation}
\label{centralheli}
\Upsilon_{L^2,transition} = 0.63650817819... + \frac{0.318899454...}
{\ln L + C} + ... \;\;.
\end{equation}
Contributions of $O(z^2)$ that we have ignored here are proportional to
$1/(\ln L + C)^2$ at the transition.

\subsubsection{The second moment correlation length}
In this section we derive a result for the dimensionless 
ratio $\xi_{2nd}/L$  analogous to eq.~(\ref{centralheli}) for the 
helicity modulus.  To this end we have to compute the XY two-point 
correlation function as a series in $z$. For the limit $L\rightarrow \infty$, 
the result can be found in the literature.  
It is important to notice that similar to the helicity modulus $O(z)$ 
contributions to the correlation function vanish.  I.e. also here 
the Gaussian result is sufficient for our purpose. The non-trivial task 
is to take properly into account the effects of periodic boundary 
conditions on the finite lattice.  The starting point of our 
calculation is the spin wave model~(\ref{winding}).  Following the 
definition~(\ref{shift}),
a difference of
variables $\tilde \phi_x$ and $\tilde \phi_y$ of the system 
with shifted boundary conditions
can be rewritten in terms of the system without shift:
\begin{equation}
 \tilde \phi_x - \tilde \phi_y = \phi_x - \phi_y +  p_1 n_1 (x_1-y_1) 
  + p_2 n_2 (x_2-y_2)
\end{equation}
with $p_i=2 \pi/L_i$. 
Using this results, the spin-spin product  can be written as
\begin{eqnarray}
\vec{s}_x \vec{s}_y 
&=& \Re \; \exp(i [\tilde \phi_x - \tilde \phi_y]) \nonumber \\
         &=& \Re \; \exp(i [\phi_x -  \phi_y])  
\exp(i  [p_1 n_1 (x_1-y_1)
            + p_2 n_2 (x_2-y_2)]) \;\;,
\end{eqnarray}
where we have interpreted $\tilde \phi_x$ as the angle of the spin $\vec{s}_x$.

The expectation value in the spin-wave limit becomes 
\begin{eqnarray}
\label{twopoint}
& & \langle  \vec{s}_x \vec{s}_y \rangle_{SW} =  \nonumber \\
& & \frac{
\sum_{n_1,n_2}   
 W(n_1,n_2) \;
\langle \exp(i [\phi_x - \phi_y]) \rangle_{0,0} 
\cos(p_1 n_1 (x_1-y_1) + p_2 n_2 (x_2-y_2))
}
{
\sum_{n_1,n_2}  
 W(n_1,n_2) } \;\;, \nonumber \\
& & \phantom{blubber}
\end{eqnarray}
where $\langle ... \rangle_{0,0}$ denotes the expectation value in a 
system with vanishing boundary shift. 
Configurations with a winding (i.e. with a shift in $\tilde \phi$)
give only minor contributions; E.g.
$W(1,0) =3.487 ... \times 10^{-6}$ for an $L^2$ lattice at $\beta=2/\pi$.

We have computed $\langle \exp(i [\phi_x - \phi_y]) \rangle_{0,0}$ 
numerically, using the lattice 
propagator. To this end, we have used lattices up to $L=2048$. For details
of this calculation see the appendix.
The results for $<s_x s_y>$
were plugged into the definition~(\ref{second}) of the second moment 
correlation length. 
Extrapolating the finite lattice results to $L\rightarrow \infty$ gives
\begin{equation}
\xi_{2nd}/L =   0.7506912... + 0.66737... \; (\beta-2/\pi)  + ... ... \;\;.
\end{equation}
Inserting $\frac{1}{\ln L + C} = \pi (\beta-2/\pi)$ 
for the critical trajectory, we obtain 
\begin{equation}
\label{xiexact}
 \xi_{2nd}/L = 0.7506912... + \frac{0.212430...}{\ln L + C}+ \;... \;\;.
\end{equation}

Note that a similar result for the exponential correlation length on a lattice
with strip geometry, i.e. an $L \times \infty$ lattice, can be found in 
the literature \cite{LuWeWo}:
\begin{equation}
\label{xiexp0}
 \xi_{exp}/L = 2 \beta \;\;.
\end{equation}
Inserting $\frac{1}{\ln L + C} = \pi (\beta-2/\pi)$ into~(\ref{xiexp0}) gives
\begin{equation}
\label{xiexp}
 \xi_{exp}/L = \frac{4}{\pi} + \frac{2}{\pi} \frac{1}{\ln L + C} +\;...
\end{equation}
at the KT-transition.
This prediction had been compared with Monte Carlo results 
in ref. \cite{me94} for lattice sizes up to $L=64$.  

It is interesting to note 
that the limit
\begin{equation}
\lim_{\xi_{exp,\infty} \rightarrow \infty} 
\left . \xi_{exp}/L \right|_{z=L/\xi_{exp,\infty}} \;\;, 
\end{equation}
where $\xi_{exp,\infty}$ is the exponential correlation length in the infinite
volume limit in the high temperature phase, 
is exactly known for any $z=L/\xi_{exp,\infty}$ \cite{Korzec03}. 
Note that this limit corresponds to the RG-trajectory that flows out of the 
point $(x,z)=(0,0)$, while the present study is concerned with the 
trajectory that flows into $(x,z)=(0,0)$.

\section{Monte Carlo Simulations}
We have simulated the XY model at $\beta=1.1199$, which is the estimate of ref. 
\cite{HaPi97} for the inverse 
transition temperature and $\beta=1.12091$ which is the 
estimate of Olsson \cite{olsson} and consistent within error-bars with
the result of Schultka and Manousakis \cite{ScMa94}. For both values of 
$\beta$,
we have simulated square lattices up to a linear lattice size of $L=2048$. 
The simulations were performed with the 
single cluster algorithm \cite{wolff}. A measurement was performed after 10 
single cluster updates. In units of these measurements, the integrated 
autocorrelation time of the magnetic susceptibility is less than one for all 
our simulations.

For each lattice size and $\beta$-value we have performed 
5.000.000 measurements, except for $L=2048$ were only 2.500.000 measurements
were performed. We have used our own implementation of the G05CAF random 
number generator of the NAG-library. For each run, 
we have discarded at least $10000$
measurements for equilibration.  Note that this is more than what is usually 
considered as safe. On a PC with an Athlon XP 2000+ CPU
the simulation of the $L=2048$ lattice at one value of $\beta$ took 
about 76 days. 

In table \ref{basic1.1199} we have summarised our results for the 
helicity modulus  $\Upsilon$, the second moment correlation length over 
the lattice size $\xi_{2nd}/L$ and the magnetic susceptibility $\chi$
at $\beta=1.1199$.  In table \ref{basic1.12091} we give analogous results 
at $\beta=1.12091$.  

\begin{table}
\caption{\sl \label{basic1.1199}
Monte Carlo results for the helicity modulus  $\Upsilon$,  
the second moment correlation length over the lattice size $\xi_{2nd}/L$ and
the magnetic susceptibility $\chi$ for two dimensional XY model on 
a square lattice of linear size $L$  at $\beta=1.1199$.
}
\begin{center}
\begin{tabular}{|r|l|l|l|}
\hline
\multicolumn{1}{|c}{$L$}&
\multicolumn{1}{|c}{$\Upsilon$} &
\multicolumn{1}{|c}{$\xi_{2nd}/L$} &
\multicolumn{1}{|c|}{$\chi$}\\
\hline
  16 & 0.72536(7)  &  0.79953(17) % 0.79801(17) %&  1.020257(17) 
                                  &\phantom{000}133.011(9) \\
  32 & 0.70883(7)  & 0.79231(18) %0.79203(18) %&  1.018684(16)  
                                  &\phantom{000}452.114(31) \\
  64 & 0.69785(7)  & 0.78701(18) % 0.78691(18) %&  1.017864(15)  
                                  &\phantom{00}1536.58(11) \\
 128 & 0.69001(7)  & 0.78310(18) % 0.78308(18) %&  1.017397(15)  
                                  &\phantom{00}5220.99(36) \\
 256 & 0.68400(7)  & 0.77977(19) %&  1.017183(14)  
                                  &\phantom{0}17729.9(1.2) \\
 512 & 0.67926(6)  & 0.77745(18) %&  1.017101(14)  
                                  &\phantom{0}60185.8(4.0) \\
1024 & 0.67544(7)  & 0.77532(19) %&  1.017037(14)  
                                  &  204160.(15.) \\
2048 & 0.67246(10) & 0.77300(28) %&  1.017022(21)  
                                  &  692146.(74.) \\
\hline
\end{tabular}
\end{center}
\end{table}

\begin{table}
\caption{\sl \label{basic1.12091}
Same as table \ref{basic1.1199} but for $\beta=1.12091$.
}
\begin{center}
\begin{tabular}{|r|l|l|l|}
\hline
\multicolumn{1}{|c}{$L$}&
\multicolumn{1}{|c}{$\Upsilon$} &
\multicolumn{1}{|c}{$\xi_{2nd}/L$} &
%\multicolumn{1}{|c}{$U_4$} &
\multicolumn{1}{|c|}{$\chi$}\\
\hline
  16 & 0.72695(7) & 0.80044(18)  % 0.79892(18)  %& 1.020155(17) 
                                  &\phantom{000}133.174(10) \\
  32 & 0.71059(7) &  0.79326(18) % 0.79287(18)  %& 1.018591(15) 
                                  &\phantom{000}452.856(31) \\
  64 & 0.69982(7) &  0.78888(18) % 0.78878(18)  %& 1.017701(15) 
                                  &\phantom{00}1540.31(11)  \\
 128 & 0.69225(7) &  0.78464(18) % 0.78462(18)  %& 1.017249(15) 
                                  &\phantom{00}5235.34(36) \\
 256 & 0.68629(7) &  0.78157(19)  %& 1.017001(14) 
                                  &\phantom{0}17794.7(1.2) \\
 512 & 0.68186(7) &  0.77951(19)  %& 1.016916(14) 
                                  &\phantom{0}60436.6(4.3) \\
1024 & 0.67826(7) &  0.77733(20)  %& 1.016862(14) 
                                  &205185.(15.) \\
2048 & 0.67528(10)&  0.77547(28)  %& 1.016798(21) 
                                  &696308.(75.) \\
\hline
\end{tabular}
\end{center}
\end{table}

First we fitted the helicity modulus $\Upsilon$ with the ansatz
\begin{equation}
\label{helian}
\Upsilon  = 0.63650817819 + const/(\ln L +C) \;\;,
\end{equation}
where  $const$ and $C$ are the free parameters of the fit.
Note that $O((\ln L)^2)$ corrections that are due to e.g. the $O(z^2)$
contribution to $\Upsilon$ 
are effectively taken into account by the fit parameter $C$.  Also
corrections \cite{AmGoGr80} 
proportional to $\ln |\ln L|/(\ln L)^2$ contribute to the value of $C$,
since $\ln |\ln L|$ varies little for the values of $L$ that enter into 
the fits.

The results of the fits 
for $\beta=1.1199$ are summarised in table \ref{fitheli1.1199}
and for $\beta=1.12091$ in table \ref{fitheli1.12091}.  For $\beta=1.1199$
the $\chi^2/$d.o.f. stays rather large even up to $L_{min}=512$.
Also the value of
$C$ is increasing steadily with increasing $L_{min}$. However this is not too
surprising, since corrections that are not taken into account in our ansatz
decrease slowly with increasing $L$. 
However, the results for $const$ approach  
the theoretical prediction $0.318899454...$ as $L_{min}$ increases.  
For $L_{min}=64$ and $128$, the $\chi^2/$d.o.f. for $\beta=1.12091$
is much larger than  for  $\beta=1.1199$. However for $L_{min}=256$ it becomes 
about one for $\beta=1.12091$. This should however be seen as a coincidence, 
since
the value of $const$ is increasing with $L_{min}$ and 
already for $L_{min}=64$ the value of $const$ is larger than the value 
predicted by the theory.  

We conclude that our fit results are consistent with $\beta=1.1199$ 
being the inverse transition temperature, while $\beta=1.12091$ is clearly 
ruled out. One should notice however that fits with ans\"atze like 
 eq.~(\ref{helian}) are problematic, since corrections that are not included
die out only very slowly as the lattice size is increased.

\begin{table}
\caption{\sl \label{fitheli1.1199} Fits of the helicity modulus at 
$\beta=1.1199$ with the ansatz~(\ref{helian}). Data with $L=L_{min}$ up to
$L=2048$ have been included into the fit.
}
\begin{center}
\begin{tabular}{|r|l|l|c|}
\hline
\multicolumn{1}{|c}{$L_{min}$}&
\multicolumn{1}{|c}{$const$} &
\multicolumn{1}{|c}{$C$} &
\multicolumn{1}{|c|}{$\chi^2$/d.o.f.}\\
\hline
 64  &  0.2957(11)  &  0.668(21) & 3.53 \\
128  &  0.2988(17)  &  0.740(37) & 2.67 \\
256  &  0.3033(29)  &  0.847(67) & 2.10 \\
512  &  0.3097(52)  &  1.01(13)  & 1.77 \\
1024 &  0.326(14)   &  1.43(37)  &  -   \\
\hline
\end{tabular}
\end{center}
\end{table}

\begin{table}
\caption{\sl \label{fitheli1.12091} Fits of the helicity modulus at 
$\beta=1.12091$ with the ansatz~(\ref{helian}). Data with $L=L_{min}$ up to
$L=2048$ have been included into the fit.
}
\begin{center}
\begin{tabular}{|r|l|l|c|}
\hline
\multicolumn{1}{|c}{ $L_{min}$}&
\multicolumn{1}{|c}{$const$} &
\multicolumn{1}{|c}{C} &
\multicolumn{1}{|c|}{$\chi^2/$d.o.f.}\\
\hline
 64     &0.3382(13) &  1.201(14) & 16.56\phantom{0} \\
128     &0.3473(21) &  1.399(42) &  9.87 \\
256     &0.3616(36) &  1.724(79) &  1.03 \\
512     &0.3688(68) &  1.90(16)  &  0.30 \\
1024    &0.377(16)  &  2.09(40)  &  - \\
\hline
\end{tabular}
\end{center}
\end{table}

Next we fitted the results for the second moment correlation length with 
an ansatz similar to that used for the helicity modulus
\begin{equation}
\label{xian}
 \xi_{2nd}/L = 0.7506912... + const/(\ln L + C) \;\;.
\end{equation}
The results of these fits are summarised in table \ref{fitxi1.1199} 
for $\beta=1.1199$ and table \ref{fitxi1.12091} for $\beta=1.12091$.  
In contrast to the helicity modulus, we get a small $\chi^2/$d.o.f. already 
for $L_{min}=64$. This might be partially due to the fact that the relative 
statistical accuracy of $\xi_{2nd}/L$ is less than that of the helicity 
modulus $\Upsilon$. The result for $const$ at $\beta=1.1199$ is quite stable
as $L_{min}$ is varied, and furthermore it is consistent with the theoretical 
prediction $const=0.212430...$ derived in this work. On the other hand, 
the fit results of $const$  at $\beta=1.12091$ are clearly larger than the 
theoretical prediction and furthermore the value of $const$ is even increasing 
as $L_{min}$ is increased. These results are consistent with the analysis 
of the helicity modulus: While our results are consistent with $\beta=1.1199$
being the inverse transition temperature,  $\beta=1.12091$ is clearly ruled 
out.

\begin{table}
\caption{\sl \label{fitxi1.1199} Fits of the  second moment correlation 
length of the lattice size $\xi_{2nd}/L$ at
$\beta=1.1199$ with the ansatz~(\ref{xian}). Data with $L=L_{min}$ up to
$L=2048$ have been included into the fit.
}
\begin{center}
\begin{tabular}{|r|l|l|c|}
\hline
\multicolumn{1}{|c}{ $L_{min}$}&
\multicolumn{1}{|c}{$const$} &
\multicolumn{1}{|c}{$C$} &
\multicolumn{1}{|c|}{$\chi^2/$d.o.f.}\\
\hline  
  64    & 0.2082(38) & 1.58(12) & 0.78 \\
 128    & 0.2086(58) & 1.59(20) & 1.03 \\
 256    & 0.2112(97) & 1.69(36) & 1.49 \\
\hline
\end{tabular}
\end{center}
\end{table}

\begin{table}
\caption{\sl \label{fitxi1.12091} Fits of the  second moment correlation 
length over the lattice size $\xi_{2nd}/L$ at
$\beta=1.12091$ with the ansatz~(\ref{xian}). Data with $L=L_{min}$ up to
$L=2048$ have been included into the fit.
}
\begin{center}
\begin{tabular}{|r|l|l|c|}
\hline
\multicolumn{1}{|c}{ $L_{min}$}&
\multicolumn{1}{|c}{$const$} &
\multicolumn{1}{|c}{$C$} &
\multicolumn{1}{|c|}{$\chi^2/$d.o.f.}\\
\hline  
  64    & 0.2435(47) & 2.26(14) & 2.24 \\
 128    & 0.2583(79) & 2.77(26) & 0.57 \\
 256    & 0.265(13)  & 3.01(46) & 0.63 \\
\hline
\end{tabular}
\end{center}
\end{table}

\subsection{The magnetic susceptibility}
The magnetic susceptibility at the transition temperature 
is predicted to behave as
\begin{equation}
\label{leadingchi}
\chi= const \; L^{2-\eta} \;\; (\ln L)^{-2 r} ... \;\;,
\end{equation}
with $r = -1/16$ and $const$ depends on the particular model. 
This result can be obtained e.g. by integration of 
\be
 \langle s_x s_y \rangle \propto R^{-1/4} (\ln R)^{1/8}
\ee
given 
in ref. \cite{AmGoGr80} for the correlation function, where $R=|x-y|$.
Leading corrections to eq.~(\ref{leadingchi}) are due to the integration
constant in eq.~(\ref{xtransition}): 
\begin{equation}
\label{integrationchi}
\chi= const \; L^{2-\eta} \;\; (\ln L + C)^{-2 r} ... \;\;.
\end{equation}

In ref. \cite{IrKe95} Irving and Kenna have simulated the same model as 
studied in this work on lattices up to $L=256$. 
Using the ansatz~(\ref{leadingchi}), leaving $r$ as free parameter, they find
$r=-0.023(10)$, which is about half of the value predicted by the 
theory. 
Later Janke \cite{Janke96} 
repeated this analysis for the XY model with the Villian
action and lattices up to $L=512$. 
He finds, also fitting with the ansatz~(\ref{leadingchi}), 
$r=-0.0270(10)$, which is consistent with the result of Irving and Kenna. 

Here we shall check whether the value of $r$ changes as larger lattice 
sizes are included into the fit. To this end, we only discuss the data 
for $\beta=1.1199$. 
In table \ref{chifit1} we give results 
for fits with the ansatz~(\ref{leadingchi}), where we have taken $-2 r$ as 
a free parameter. The $\chi^2/$d.o.f. is very large up to $L_{min}=256$.
For $L_{min}=32$ our results for $-2 r$ is slightly larger than that 
of refs. \cite{IrKe95,Janke96}. As we increase $L_{min}$ also $-2 r$ increases.
However, even for $L_{min}=512$, the result for $-2 r$ is by more than 70 
standard deviations smaller than the value predicted by the KT-theory.

\begin{table}
\caption{\sl \label{chifit1} Fits of the magnetic susceptibility at 
$\beta=1.1199$ with the ansatz~(\ref{leadingchi}). 
 Data with $L=L_{min}$ up to
$L=2048$ have been included into the fit.
}
\begin{center}
\begin{tabular}{|r|l|l|c|}
\hline
\multicolumn{1}{|c}{ $L_{min}$}&
\multicolumn{1}{|c}{$const$} &
\multicolumn{1}{|c}{$-2 r$} &
\multicolumn{1}{|c|}{$\chi^2/$d.o.f.}\\
\hline
  32  &   0.9611(2) & 0.0699(1)   & 382.5 \\
  64  &   0.9539(3) & 0.0741(2)   & 119.2 \\
 128  &   0.9485(4) & 0.0772(2)   &  \phantom{0}35.7 \\
 256  &   0.9439(6) & 0.0798(3)   &  \phantom{00}5.2 \\
 512  &   0.9412(11)& 0.0812(6)   &  \phantom{00}1.5 \\
\hline
\end{tabular}
\end{center}
\end{table}

Next we checked whether this apparent discrepancy can be resolved by
adding the leading correction predicted by the theory as free parameter
to the fit. In table \ref{chifit2} we give our results for fits with 
the ansatz~(\ref{integrationchi}), where we have fixed $-2 r=1/8$.  We see
that already for $L_{min}=128$ an acceptable $\chi^2/$d.o.f. is reached.

\begin{table}
\caption{\sl \label{chifit2} Fits of the magnetic susceptibility at
$\beta=1.1199$ with the ansatz~(\ref{integrationchi}), 
fixing the exponent to the value $-2 r=1/8$.
Data with $L=L_{min}$ up to
$L=2048$ have been included into the fit.
}
\begin{center}
\begin{tabular}{|r|l|l|c|}
\hline
\multicolumn{1}{|c}{ $L_{min}$}&
\multicolumn{1}{|c}{$const$} &
\multicolumn{1}{|c}{$C$} &
\multicolumn{1}{|c|}{$\chi^2/$d.o.f.}\\
\hline
   8  &   0.8121(1) & 4.423(9)   &  307.2 \\
  16  &   0.8146(1) & 4.187(11)  &  115.0 \\
  32  &   0.8170(2) & 3.953(14)  &  \phantom{0}32.5 \\
  64  &   0.8187(2) & 3.786(20)  &  \phantom{00}6.6 \\
 128  &   0.8197(3) & 3.690(28)  &  \phantom{00}1.5 \\
 256  &   0.8204(5) & 3.625(43)  &  \phantom{00}0.4 \\
\hline
\end{tabular}
\end{center}
\end{table}

Finally we performed fits with the ansatz~(\ref{integrationchi}), where 
now also $-2 r$ is used as free parameter. The results are summarised in table
\ref{chifit3}.  The $\chi^2/$d.o.f. becomes acceptable for $L_{min}$
starting from $L_{min}=128$.  Now the fit results for $-2 r$ for $L_{min}=128$
and $256$ are consistent within the statistical errors 
with the theoretical prediction.

\begin{table}
\caption{\sl \label{chifit3} Fits of the magnetic susceptibility at
$\beta=1.1199$ with the ansatz~(\ref{integrationchi}).
 Data with $L=L_{min}$ up to
$L=2048$ have been included into the fit.
}
\begin{center}
\begin{tabular}{|r|l|l|l|c|}
\hline
\multicolumn{1}{|c}{ $L_{min}$}&
\multicolumn{1}{|c}{$const$} &
\multicolumn{1}{|c}{$C$} &
\multicolumn{1}{|c}{$-2 r$} &
\multicolumn{1}{|c|}{$\chi^2/$d.o.f.}\\
\hline
  32  &  0.685(15) & 7.73(45) & 0.177(6)  &  4.92 \\
  64  &  0.747(19) & 5.83(55) & 0.152(7)  &  1.97 \\
 128  &  0.789(26) & 4.58(76) & 0.136(10) &  1.49 \\
 256  &  0.857(38) & 2.5(1.1) & 0.112(14) &  0.01 \\
\hline
\end{tabular}
\end{center}
\end{table}

We conclude that the apparent discrepancy with the KT-theory that was
observed in refs. \cite{IrKe95,Janke96} can be resolved by adding a correction 
term, which is predicted by the KT-theory, to eq.~(\ref{leadingchi}).

\section{Summary and Conclusions}
We have studied the finite size behaviour of various quantities at  
the Kosterlitz-Thouless transition of the two-dimensional 
XY model.  For the helicity modulus $\Upsilon$ the value 
at the Kosterlitz-Thouless transition 
in the $L\rightarrow \infty$ limit and the leading 
logarithmic corrections to it are exactly known. 
Here, we have derived the analogous result~(\ref{xiexact}) for  
the second moment correlation length over the lattice size $\xi_{2nd}/L$:
\begin{eqnarray}
 \xi_{2nd}/L = 0.7506912... + \frac{0.212430...}{\ln L + C}+ \;... \;\;.
\nonumber
\end{eqnarray}
We have performed
Monte Carlo simulations of the 2D XY model at $\beta=1.1199$ and 
$\beta=1.12091$, which are the estimates of the transition temperature
of ref. \cite{HaPi97} and ref. \cite{olsson}, respectively. 
Using the single cluster algorithm we simulated lattices of a size
up to $2048^2$, which is by a factor of $5^2$ 
larger than the lattices that
had been studied in ref. \cite{olsson}.  
Analysing our data for the helicity modulus $\Upsilon$ and the ratio 
$\xi_{2nd}/L$ we confirm $\beta=1.1199$ as transition temperature, while 
$\beta=1.12091$ is clearly ruled out. 

Fitting Monte Carlo data with the ans\"atze~(\ref{helian},\ref{xian}) is 
certainly a reasonable method to locate the transition temperature and 
to verify the Kosterlitz-Thouless nature of the transition.
However one should note that
the large values of $\chi/$d.o.f.
of our fits and the running of the fit parameter $C$ with the smallest 
lattice size $L_{min}$ that is included into the fits, 
indicate that subleading corrections that are not taken into account in the 
ans\"atze~(\ref{helian},\ref{xian})
are still large for the lattice sizes that we have studied.
Since these corrections decay only logarithmically with the lattice size, 
it is difficult to estimate the systematic errors that are due to these 
corrections. 

Finally we studied the finite size scaling 
of the magnetic susceptibility.  At the transition it should behave 
like $\chi \propto L^{2-\eta} \ln L^{-2 r}$ with $\eta=1/4$ and $r=-1/16$.
However, fitting numerical data, the authors of 
refs. \cite{IrKe95,Janke96} found $r=-0.023(10)$ and $r=-0.0270(10)$, 
respectively. Including larger lattices into the fits, our result for $r$ moves
toward the predicted value. Extending the ansatz to 
$\chi \propto L^{2-\eta} (\ln L+C)^{-2 r}$, where $C$ is an additional free 
parameter consistent with the theory, the apparent contradiction is completely
resolved: For a minimal lattice size $L_{min}= 256$ that is included 
into the fit, we get $r=-0.056(7)$.

\section{Acknowledgement}
I like to thank Ettore Vicari for discussions and a critical reading of 
the draft. Some of the simulations have been performed
at the Humboldt-Universit\"at zu Berlin,
where I stayed as guest in October 2004. I like to thank 
Vincenzo Alba for spotting  the error in $\xi/L$ given in 
tables 1 and 2 that has been corrected in the present version of the paper.

\section{Appendix: The correlation function at \\ $z=0$}
Here we compute the spin-spin correlation function for $z=0$, i.e. for the 
spin wave approximation, for finite lattices with periodic boundary 
conditions.

To this end  let us first summarise a few basic formula on multi-dimensional
Gaussian integrals as they can be found in text books on field theory.

Our starting point is the generating functional
\be
\label{master}
\frac{1}{Z} \int D [\phi]  
 \exp\left(-\frac{1}{2 \beta} (\phi, A \phi) + i k \phi\right)
 = \exp\left(-\frac{\beta}{2} (k,A^{-1} k) \right)
\ee
where 
\be
\frac{1}{2 \beta} (\phi, A \phi)=
\frac{1}{2 \beta} \sum_{x,y} A_{xy} \phi_x \phi_y =
\frac{1}{2 \beta} 
\sum_{x,\mu} \left [(\phi_x - \phi_{x+\hat \mu})^2 + m^2 \phi_x^2 \right]
\ee
is the action of the Gaussian model on a square lattice and  the partition 
function is given by
\be
 Z =  \int D [\phi] \exp\left(-\frac{1}{2 \beta} (\phi, A \phi) \right) \;\;
\ee
with
\be
 \int D [\phi] = \prod_x \int \mbox{d} \phi_x \;\;.
\ee
For a square lattice with periodic boundary conditions $A^{-1}$ can 
be easily obtained using a Fourier transformation:
\ba
(A^{-1})_{xy} &=& \frac1{L^2} \sum_{p}
\frac{\mye^{ip(x-y)}}{\hat p^2 + m^2} \, , \nonumber \\
\hat p^2 &=& 4 - 2 \cos p_1 - 2 \cos p_2 \, ,
\ea
where the $p_i$, $i=1,2$
are summed over the values $\{0,...,L-1\} \cdot(2\pi/L)$.
Here we are interested in the massless limit $m \rightarrow 0$. 
Note that for $\sum_x k_x = 0$ 
the contributions to $(k,A^{-1} k)$ from $(p_1,p_2)=(0,0)$ exactly cancel,
while for 
$\sum_x k_x \ne 0$, in the limit $m \rightarrow 0$, 
the right hand side of eq.~(\ref{master}) 
vanishes due to the divergent zero-momentum contributions to $(k,A^{-1} k)$. 
Hence we get:
\ba
\label{masterfinal}
&& \lim_{m \rightarrow 0}  \frac{1}{Z} \int D [\phi]
 \exp\left(-\frac{1}{2 \beta} (\phi, A \phi) + i k \phi\right)  \nonumber \\
&& \phantom{000000000000000000000} =
\left\{ \begin{array}{ll}
        \exp \left[ - \half \beta (k,C k) \right]
        & \mbox{, if } \sum_x k_x = 0 \\
%%c else zu otherwise
        0 &\mbox{, otherwise .} \\
        \end{array}
\right.
\ea
with
\be
\label{Cxx}
C_{xy} = \frac1{L^2} \sum_{p \ne 0}
\frac{\mye^{ip(x-y)}-1}{\hat p^2} \;\;. \\
\ee
Note that adding a constant to $C_{xy}$ does not change the result. Here
we have chosen this constant such that $C_{xx}=0$. 

Now we are in the position to compute the 
two-point correlation function~(\ref{twopoint})
required for the computation of the second moment correlation 
length~(\ref{second}):
\begin{equation}
 \langle \exp(i [\phi_x-\phi_y]) \rangle_{00}
 = \exp \left[ \beta C_{xy} \right] \;\;.
\end{equation}

Due to translational invariance, it is sufficient to compute $g(x)=C_{(0,0),x}$,
for all lattice sites $x$. Employing the reflection symmetry of the lattice 
with respect to various axis the number of sites can be further reduced by
a constant factor. Still, the direct implementation of eq.~(\ref{Cxx}) 
would
results in a computational effort $\propto V^2$ for the calculation 
of $\xi_{2nd}$, where $V$ is the number of lattice points. A more efficient 
method is discussed below.

First we compute $g(x)$ with $x=(x_1,0)$ for $x_1>0$:
\begin{equation}
\label{super}
 g(x_1,0) = \frac1{L^2} \sum_{p_1 \ne 0} \; Q(p_1) \;[\mye^{i p_1 x_1}-1]
\end{equation}
with
\begin{equation}
 Q(p_1) = \sum_{p_2} \frac{1}{\hat p^2} \;\;.
\end{equation}
I.e. these $g(x)$ can be computed with an effort proportional to $V$.

Next we notice that $g(x)$ satisfies 
Poisson's equation (see e.g. ref. \cite{LuWe95} and refs. therein):
\ba
\label{poisson}
  4 g(x)-g(x-(1,0))-g(x+(1,0))-g(x-(0,1))- g(x+(0,1)) &=& \nonumber \\
  \frac1{L^2} \sum_{p \ne 0} 
 \frac{\mye^{ipx} 
(4-\mye^{ip_1}-\mye^{-ip_1} -\mye^{ip_2}-\mye^{-ip_2})}{\hat p^2} = 
\frac1{L^2} \sum_{p \ne 0}\frac{\mye^{ipx} \hat p^2}{\hat p^2} = 
\phantom{xxxxxxx} & &
 \nonumber \\
  \frac1{L^2} \sum_{p \ne 0} \mye^{ipx} =  
\left\{ \begin{array}{ll}
       1-L^{-2} & \mbox{, if } x  = (0,0) \\
%%c else zu otherwise
        -L^{-2} &\mbox{, otherwise .} \\
        \end{array}
\right.  \phantom{xxxxxxxxxxxxxxxxxx} &&
\ea
In principle, the remaining $g(x)$ can now be computed recursively, using 
eq.~(\ref{poisson}).  First one has to note that 
$g(x_1,1)=g(x_1,-1)$, where we identify $L-1$ with $-1$, for symmetry 
reason. Therefore
\begin{equation}
g(x_1,1)=\frac{1}{2} [4 g(x_1,0)- g(x_1-1,0) - g(x_1+1,0) + L^{-2}] \; .
\end{equation}
Then for
$x_2>1$ one gets
\begin{eqnarray}
g(x_1,x_2)= 4 g(x_1,x_2-1) \phantom{xxxxxxxxxxxxxxxxxxxxxxxxxxxxxxxxx}
 \nonumber \\
              - g(x_1-1,x_2-1)-g(x_1+1,x_2-1)-g(x_1,x_2-2) +L^{-2}\;.
\end{eqnarray}
Unfortunately, rounding errors rapidly accumulate, and the recursion is 
useless, at least when using double precision floating point numbers, for the 
lattice sizes we are aiming at.

Instead, we have used an iterative solver to solve eq.~(\ref{poisson}).
We imposed $g(x_1,0)=g(0,x_1)$ obtained from eq.~(\ref{super}) as 
Dirichlet boundary conditions.  As solver we have 
used a successive overrelaxation (SOR)
algorithm. With the optimal overrelaxation parameter, the
computational effort is proportional to $L^3$. We controlled the numerical
accuracy of the solution by computing $g(x)$ from eq.~(\ref{Cxx}) for a few
distances $x$. 
Since we could extract sufficiently accurate results for the limit
$L \rightarrow \infty$ from lattice sizes up to 
$L=2048$, we did not implement more advanced solvers like e.g. 
multigrid solvers.

\end{document}